\documentclass[twocolumn,prb,aps,longbibliography,superscriptaddress,floatfix,tightenlines]{revtex4-2}
\usepackage{lipsum}
\usepackage{graphicx}
\usepackage{dcolumn}
\usepackage{bm}
\usepackage[mathlines]{lineno}
\usepackage{graphicx}
\usepackage{subfigure}
\usepackage{physics}
\usepackage{amsmath}
\usepackage{times}
\usepackage{amssymb}
\usepackage{mathrsfs}
\usepackage{amsfonts}
\usepackage[T1]{fontenc}
\usepackage[dvips]{color}
\usepackage[colorlinks,bookmarks=false,citecolor=blue,linkcolor=magenta,urlcolor=blue]{hyperref}

\begin{document}

\preprint{APS/123-QED}

\title{{\color{black}Discrete time crystals} in one-dimensional classical
  Floquet systems with nearest-neighbor interactions
}

\author{Zhuo-Yi Li}
\affiliation{School of Physics and Optoelectronics, South China
University of Technology, Guangzhou 510640, China}

\author{Yu-Ran Zhang}
\email{yuranzhang@scut.edu.cn}
\affiliation{School of Physics and Optoelectronics, South China
University of Technology, Guangzhou 510640, China}

\date{\today}

\begin{abstract}
  Prethermal discrete time crystals (PDTCs), an emergent
  non-equilibrium phase of matter, have been studied
  in two- and higher-dimensional lattices with nearest-neighbor (NN)
  interactions and
  one-dimensional (1D) lattices with long-range interactions.
  However, different from prethermalization that
  can be observed in 1D Floquet classical spin systems with NN
  interactions, classical PDTCs in Floquet 1D  systems with only
  NN interactions have not been proposed before.
  Here, we demonstrate the emergence of disorder-free
  {\color{black}discrete time crystals (DTCs)} in 1D Floquet classic spin
  systems with NN interactions. We show that the thermalization time
  {\color{black}first} grows
  exponentially as the driving frequency increases and {\color{black}
  is then saturated, which} depends on the energy density of the initial state.
  {\color{black}Since thermalization of the
  effective Hamiltonian is slow, there is no typical prethermalization
  and PDTCs in the Floquet system before final thermalization}.
  The robustness of DTC order
  is verified by introducing imperfect spin flip operations.
  Our work provides an exploration of quantum characteristics, when
  considering the classical counterparts of
  quantum phenomena, and will be helpful for further investigations
  of both classical and quantum prethermal systems
  and discrete time-crystalline order.
\end{abstract}

\maketitle


\section{Introduction}
Eigenstate thermalization hypothesis (ETH) posits that the unitary
evolutions of quantum many-body systems
can eventuate in an equilibrium pure state described by statistical
mechanics~\cite{Deutsch1991, Srednicki1994,
Srednicki1999,Rigol2008, DAlessio2016}.
However, there exist several violations of ETH. A straightforward
case is the quantum systems with integrability, which
do not display an ergodic behavior~\cite{Kinoshita2006, Rigol2007,
Gong2022}. For nonintegrable systems,
thermalization can be avoided by introducing disorder, which is
called many-body localization (MBL)~\cite{
  Basko2006, Pal2010, Kjaell2014,Huse2014, Serbyn2016,
  Nandkishore2015,Abanin2019,Abanin2017b,
Alet2018,Sierant2022,Schulz2019,Vosk2015}. For an integrable system
or a nonintegrable system with strong disorder, most initial states can
violate ETH, which are called the strong violations of ETH.
Moreover, there are also special eigenstates violating ETH in some
nonintegrable and disorder-free systems, e.g., the emergence of
quantum many-body scarred
states~\cite{Bernien2017,Turner2018,Choi2019,Lin2020,Su2023,Ge2024}.
For a much smaller amount of quantum many-body scarred states
compared to the Hilbert space dimension, they are considered
weak violations of ETH.
Prethermalization, another type of weak violations of ETH,  results
in a metastable plateau of the non-equilibrium dynamics of the energy
density. Concretely, the systems experience 
an approximate thermalization in terms of an effective static Hamiltonian in
the prethermal regime before their ultimate
thermalizing~\cite{Berges2004, Mori2016,Kuwahara2016, Else2017,
  Machado2020,
  Luitz2020,Ho2023,Zhao2021,He2023,Yang2023,Jin2023,Yan2024,McRoberts2023,Fu2024,
Abanin2017}.
In the Floquet prethermal regime, the stroboscopically measured
observables reach a plateau under periodic driving before the final
thermalization.
In contrast to MBL, the thermalization time of prethermalization,
defined as the end of the prethermal plateau,
grows exponentially with the increase of the driving frequency.
Moreover, the energy density of the initial state also
influences the rate of Floquet heating, which is also characterized
by the thermalization time~\cite{Abanin2017, Canovi2016,
  Weidinger2017, Abanin2017a,
Mallayya2019}.

Discrete time crystals (DTCs) exhibit rigid subharmonic oscillations
that result from a combination of many-body
interactions, collective synchronization, and ergodicity
breaking~\cite{Zaletel2023,Yousefjani2025}.
MBL and prethermalization are two distinct mechanisms for stabilizing DTC order.
Prethermal discrete time crystals (PDTCs), an emergent
out-of-equilibrium phase, occur in the prethermal
regime, where thermalization is delayed exponentially in the driving
frequency. The PDTCs have been theoretically investigated and
experimentally observed in periodically and quasi-periodically
driving quantum systems~\cite{Zeng2017, Mizuta2019, Khemani2019,
  Rovny2018, Yao2017, Yao2018,
  Beatrez2023,Ying2022,Chen2024,Yue2022,Liu2024,Yang2024,
DeNova2022,Zhuang2021,He2024,Liu2025}.
In addition, the PDTCs have been proposed in classical
systems~\cite{Rajak2018, Mori2018,  Rajak2019,Howell2019,
  Gambetta2019, Khasseh2019, Heugel2019, Yao2020, Malz2021,Pizzi2021b,
Else2017,Else2020,Machado2020,Khemani2017,Cosme2023,Gallone2024}.
For closed quantum Floquet systems, PDTCs have only been probed in
two-dimensional (2D) and three-dimensional (3D) systems with
nearest-neighbor (NN) interactions~\cite{Else2017} and one-dimensional (1D)
systems with long-range interactions~\cite{Machado2020}. Several work
show that quasiperiodic driving
can lead to multiple time-translation symmetries~\cite{Else2020},
which can be used to induce PDTCs in 1D quantum systems
with NN interactions~\cite{Zhao2021}.
Moreover, PDTCs have also
been proposed in classical Floquet systems in 2D and 3D lattices with
NN interactions and in 1D lattice with long-range
interactions~\cite{Pizzi2021,Pizzi2021a,Ye2021}. Although
prethermalization can be observed in 1D
classical Floquet systems with NN interactions~\cite{Howell2019},
PDTCs in 1D classical systems with only NN interactions have
never been proposed before.

In this paper, we demonstrate the emergence of {\color{black} DTCs} in 1D
classical Floquet spin systems with only NN interactions. We show that the
{\color{black}prethermalization-like DTCs} have a thermalization
time that {\color{black}first} grows exponentially with the increase of
the driving frequency {\color{black}and then reaches saturation for a large
driving frequency.} With different initial states, the dependence of
thermalization timescales on the initial energy density is shown.
By implementing the imperfect spin flip operations, we demonstrate
the robustness of the DTC order in 1D classical systems with NN interactions.
By leveraging the interplay between nonlocal correlations and
emergent symmetries to overcome the limitations of classical
short-range systems, our work extends the understanding of classical
DTCs and will be helpful
for further studies of PDTCs in both classical and quantum systems.

\begin{figure}[t]
  \includegraphics[width=0.48\textwidth]{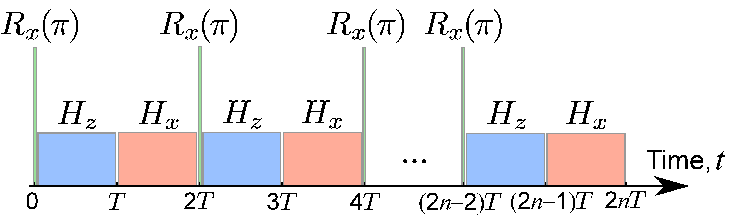}
  \caption{\label{fig:pulses}
    Schematic diagram for the Floquet driving system. During the
    first half period and second half period, the dynamics of the
    system is governed by $H_z$ and $H_x$, respectively.
    At the beginning of each period, a global spin flip operation
  $R_x(\pi)$ along the $x$-axis is applied.}
\end{figure}

\section{Hamiltonian and initial states}
We consider a 1D Floquet system consisting of two different classical
spin models of $N$ classical spins with NN interactions and local
fields. The Floquet system
possesses a period of $2T$ with a Hamiltonian:
\begin{equation}\label{eq1}
  H(t) = \left\{
    \begin{aligned}
      &H_z,\ \text{for }t\ \text{mod}\ T \in[0, T] \\
      &H_x,\ \text{for }t\ \text{mod}\ T \in[T, 2T] ,
    \end{aligned}
    \right.
  \end{equation}
  where
  \begin{align}
    H_z&=\sum_{i=1}^N (4J_z S_{i}^z S_{i+1}^z + 2 b_z S_{i}^z),\label{eq2}\\
    H_x&=\sum_{i=1}^N (4J_x S_{i}^x S_{i+1}^x + 2  b_x S_{i}^x),\label{eq3}
  \end{align}
  with $J_{z,x}$ being real interaction strengths and $b_{z,x}$ being
  the real local fields along the $z$- and $x$-axes, respectively.
  For the implementation of the {\color{black}DTCs}, we apply a global $x$-flip
  operation $R_x(\pi)$ at the beginning
  of each driving period, see Fig.~\ref{fig:pulses}.

  The initial directions of the classic spin vectors are randomly
  chosen from a unit sphere,
  whose polar angles obey a Gaussian distribution with a mean $\pi$
  and a standard deviation $2\pi W$,
  and the azimuth angle forms an uniform distribution between $0$ and
  $2\pi$. In the following context,
  we choose $W=0.1$, which describes that the classical spin system
  is initially at a finite temperature. Furthermore, the
  extension of this model described with Eqs.~(\ref{eq1}--\ref{eq3})
  can be applied to study other out-of-equilibrium phases of matter
  induced by random multipolar drivings, e.g., time rondeau crystals
  by introducing the staggered
  driving amplitudes as $b_x\to(b_x\pm\delta)$~\cite{Zhao2021,Moon2024,Liu2025}.

  \section{\label{section:dynamics}Time evolutions of Floquet system
  and observables}
  To implement Floquet driving, we choose two different Hamiltonians
  in Eqs.~(\ref{eq2},\ref{eq3}), dominating the time evolution of two
  half periods $2T$, respectively, see Fig.~\ref{fig:pulses}.
  In each half period $T$, the motion of the system is described by
  the {\color{black}equations of motion}:
  \begin{equation}
    \frac{\partial}{\partial t}S_i^\alpha = \{S_i^\alpha, H_{x,z}(t) \},
  \end{equation}
  with $\{\circ  ,\circ \}$ being the Poisson bracket as
  \begin{equation}
    \{f,g\}=\sum_{i=1}^N \left( \frac{\partial f}{\partial
      q_i}\frac{\partial g}{\partial p_i} - \frac{\partial f}{\partial
    p_i}\frac{\partial g}{\partial q_i} \right).
  \end{equation}
  Here, $S_i^\alpha$ for $\alpha\in \{x,y,z\}$ are classical spin
  components at the $i$-th site, and $q_i$ and $p_i$ are coordinates
  in phase space. Here, we use
  \cite{Howell2019,Pizzi2021,Pizzi2021a}
  \begin{equation}
    \{S_i^\alpha, S_j^\beta\} \equiv
    \delta_{ij}\epsilon_{\alpha\beta\gamma}S_i^{\gamma},
  \end{equation}
  which keeps the length of all classical spin vectors unchanged
  during the time evolution,
  see Appendix~\ref{app.unitary} for more details.

  As illustrated in Fig.~\ref{fig:pulses},
  during each driving period, the motion of the system is equivalent
  to applying local rotations on classical spins
  along the $z$-axis and then the $x$-axis, following a global
  flipping operation:
  \begin{equation}
    \mathbf{S}_i \to R_{x}(\theta_{i}^x) R_{z}(\theta_{i}^z) R_x(\pi)
    \mathbf{S}_i,
  \end{equation}
  where $\mathbf{S}_i$ denotes the spin vector at the $i$-th site,
  $R_\alpha(\theta_{i}^\alpha)$ denotes the rotation operation on the
  local spin at the $i$-th site
  with an angle $\theta_{i}^\alpha$ along the $\alpha$-axis for
  $\alpha = x,z$, and $R_x(\pi)$ denotes the global flipping
  operation on all spins along the $x$-axis.
  Here, we have $\theta_{i}^\alpha = (4 J_\alpha \bar{S}^\alpha_{i} +
  2 b_\alpha)T$, and
  $\bar{S}^\alpha_{i}=S^\alpha_{i-1}+S^\alpha_{i+1}$~\cite{Howell2019}.

  To observe the dynamical prethermal phenomenon,
  we first consider the energy absorption of the Floquet system,
  indicated by the average energy density of
  one period $2T$. We consider the
  effective Hamiltonian to the zeroth order~\cite{Howell2019} and
  check the average energy density:
  \begin{equation}
    \overline{H}_{\textrm{eff}} \equiv \frac{1}{N}\sum_{i=1}^N
    (2 J_z S_{i}^z S_{i+1}^z + 2 J_x S_{i}^x S_{i+1}^x  + b_z S_{i}^z
    + b_x S_{i}^x),
  \end{equation}
  which is divided by the system size $N$ (the length of the
  classical spin chain in our model).
  However, the prethermal phenomenon, identified by the exponentially
  suppressed energy absorption from the
  dynamics of $\overline{H}_{\mathrm{eff}}$, does not sufficiently
  lead to the presence of PDTCs, which have been shown
  in previous studies~\cite{Ye2021, Pizzi2021a}.

  Then, to observe subharmonic response to determine DTC order of
  this Floquet system,
  we also investigate the magnetization along $z$ direction:
  \begin{equation}
    {M}^z \equiv \frac{1}{N}\sum_{i=1}^N S_i^z,
  \end{equation}
  which is the mean $z$-component of spin vectors.
  Finally, we consider the normalized decorrelator
  \begin{equation}
    d \equiv \frac{1}{d_f}\sqrt{\frac{1}{N}\sum_i |
    \mathbf{S}_i-\mathbf{S}_i^\prime | ^2},
  \end{equation}
  to characterize the final ergodic behavior for retrieving the
  thermalization time $\tau^*$ of the Floquet system.
  Here, $\mathbf{S}_i^\prime$
  is an initially perturbed version of the original system's
  classical spin vector at $i$-th site, with a deviation of
  $2\pi \Delta\delta_i$ in both polar and azimuth angles, where
  $\Delta=0.01$, and all $\{\delta_i\}$ follow
  a Gaussian distribution. The normalization factor $d_f$ is chosen
  as $\sqrt 2$, which is the statistical average
  of the decorrelator, when the system finally thermalizes to
  infinite-temperature states~\cite{Pizzi2021}.


  \begin{figure}[t]
    \includegraphics[width=0.48\textwidth]{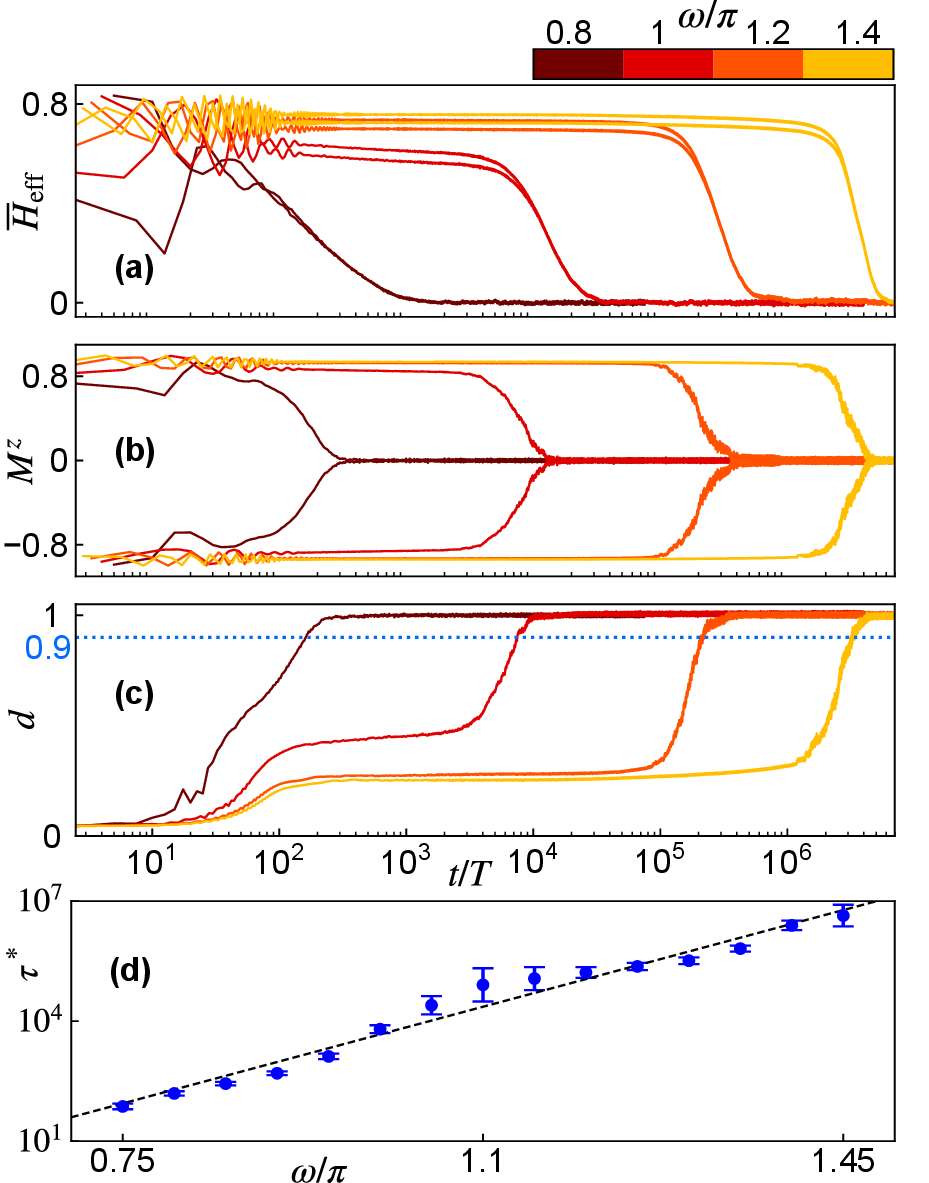}
    \caption{\label{fig:1dNN-dmH}Characterization of {\color{black}
      prethermalization-like DTCs} in 1D
      classical systems with NN interactions.
      (a--c) Dynamics of the average energy density
      $\overline{H}_\mathrm{eff}$ (a), the magnetization along
      $z$-direction $M^z$ (b), and the normalized decorrelator $d$
      (c)  versus the evolution time $t/T$, for
      different driving frequencies $\omega$. Here, $J_z=0.399$,
      $J_x=0.011$, $b_z=-0.016$, $b_x=-0.3$,
      $N=100$, and the periodic boundary conditions (PBCs) are considered.
      {\color{black}For each driving frequency, two curves in the
        same color in (a,~b) denote values at even
      and odd periods, respectively}.
      The driving frequency $\omega$ ranges from $0.8\pi$ to
      $1.4\pi$. (d) Average thermalization time $\tau^{*}$
      versus the driving frequency $\omega$, which is obtained as the
      time when $d$ reaches $0.9$, as
      illustrated in (c). The mean values shown with  errorbars for
      the one standard derivation (1SD) are
      obtained with 100 random initial states. The black dashed line
      denotes the exponential fitting
      $e^{c\omega}$.
      {\color{black}Hereafter, the same set of parameters are used
      unless stated otherwise.}
    }
  \end{figure}

  {As a weak violation of ETH, neither prethermalization nor PDTCs
    can evade final thermalization,
    and the thermalization time is exponentially long with the
    increase of the frequency of Floquet driving.
    Before final thermalization, the time evolution of the Floquet
  system with a Hamiltonian $H(t)$} at
  stroboscopic times {\color{black}(in the toggling frame)} can be
  approximated by considering the
  evolution with a time-independent Hamiltonian, $D$, up to an
  accuracy {\color{black} that is exponentially
  small in $1/T$}~\cite{Else2017}. For the PDTCs, {\color{black}$D$
  spontaneously
  breaks the spin-flip symmetry compared to prethermalization, where
  no spin-flips are considered, while both of them share exponentially
  postponed thermalization.}
  Because of the fast thermalizing dynamics of $D$, the
  expectation values of local observables at even multiples of the
  period are approximately the same as
  the equilibrium value of $D$, giving rise to the appearance of
  prethermal regime.
  {\color{black}The expectation values of local observables at odd
    multiples of the period are not the same
    with those at even periods. This is because of the
    initial-state-dependent spontaneously
    symmetry-breaking~\cite{Else2017, Machado2020}, which leads to a
  subharmonic response.}
  After an exponentially long period of time, accumulated error
  invalidates the approximation, and
  the Floquet system thermalizes~\cite{Else2017}. Here, we
  investigate the thermalization time $\tau^*$
  for different driving frequencies $\omega$, which is defined by the
  time, when the normalized
  decorrelator $d$ exceeds $0.9$, see the horizontal blue dotted line
  in Fig.~\ref{fig:1dNN-dmH}(c).
  For $t>\tau^*$, the prethermal plateau of $d$ is considered to
  vanish, which starts when $D$
  thermalizes fast~\cite{Else2017,Pizzi2021, Pizzi2021a}.

  Hereafter, the mean values of the three observables introduced
  above are calculated from 100
  random initial states at the stroboscopic times $t=mT$, with $m\in\mathbb{N}$.
  The thermalization times $\tau^*$ for different driving frequencies
  $\omega$ are also obtained
  from the dynamics of the Floquet system starting with 100 random
  initial states.

  {\color{black}
  \section{Identification of prethermalization-like discrete time crystals}
  }

  The numerically calculated mean values of the energy density
  $\overline{H}_\mathrm{eff}$, the
  magnetization $M^z$, and the normalized decorrelator $d$ versus
  different driving frequencies,
  $\omega/\pi=0.8$, 1, 1.2, and $1.4$, are shown in
  Fig.~\ref{fig:1dNN-dmH}(a--c).
  The numerics of the average energy density $\overline{H}_\mathrm{eff}$ in
    Fig~\ref{fig:1dNN-dmH}(a) show that the energy absorption rate
    decreases exponentially with the increase of the driving frequency, as
    discussed in Ref.~\cite{Pizzi2021a}. This indicates {\color{black}one
    feature of prethermalization and PDTCs.}
  For the normalized decorrelator  $d$
  as shown in Fig.~\ref{fig:1dNN-dmH}(c), the time when $d$
  approximates to the infinite-temperature value,
  $d=1$, is also exponentially postponed as the driving frequency increases.

  Different from the model investigated in Ref.~\cite{Pizzi2021a}, we
  demonstrate the emergence of the {\color{black}DTCs} in 1D classical systems
  with only NN interactions from the results of
  $M^z$, as shown in Fig.~\ref{fig:1dNN-dmH}(b).
  Before the thermalization time $\tau^*$, the $M^z$ oscillates
  between the positive and negative values
  at stroboscopic measurements of each period, which means that the
  disorder-free classical Floquet
  spin system spontaneously breaks the discrete time translation symmetry of the
  system, and obviously, DTC order appears.

  \begin{figure}[t]
    \includegraphics[width=0.48\textwidth]{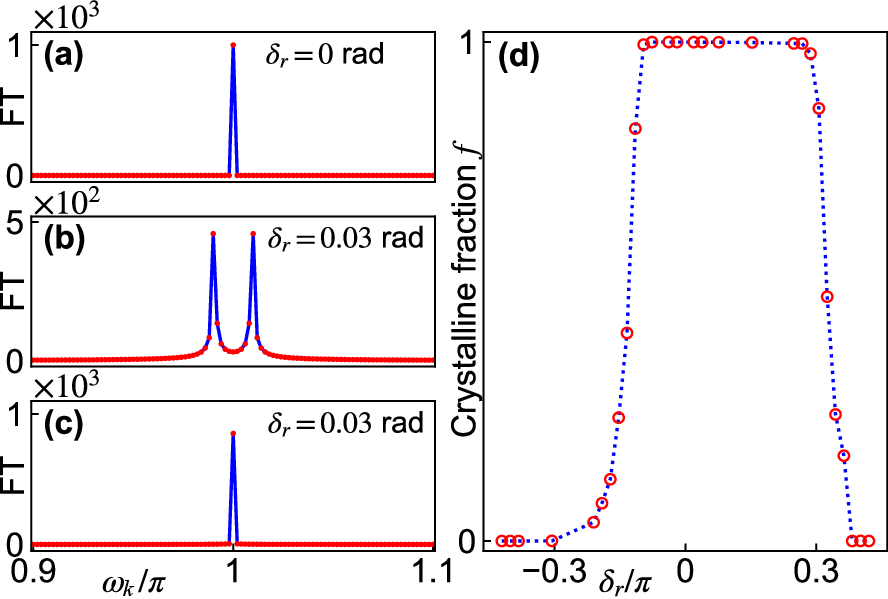}
    \caption{\label{fig:DFT}Robustness of {\color{black}DTCs} in 1D classical
      systems with NN interactions by
      introducing an imperfect global $x$-flip operation with a small
      error $\delta_r$. (a--c) Fourier transform
      (FT) signals of the average magnetization $M^z$ (c)  are
      compared with those of the trivial driving
      without any interaction (a) and (b) for $\delta_r=0$~rad and
      0.03~rad, respectively, with the driving
      frequency $\omega$ being set as $\pi$. The first 500 periods of
      $M_z$ are chosen to calculate the FT
      signals, with an accuracy of $10^{-3}$~rad in the spectrum. For
      $\delta_r=0.03$~rad, robust DTC order is
      observed in (c) with one peak of the FT signals, while the FT
      signals of the trivial imperfect global $x$-flip
      in (b) splits into 2 peaks. (d) Crystalline fraction $f$ as a
      function of $\delta_r$. The crystalline fraction
      $f$ approaching 1 indicates the robust subharmonic response of PDTC order.
      The subharmonic response is stable for $\delta_r / \pi
      \in[-0.1,0.27]$, with a maximum tolerance of
    the error of around $\pi/10$.}
  \end{figure}

  \begin{figure}[t]
    \includegraphics[width=0.48\textwidth]{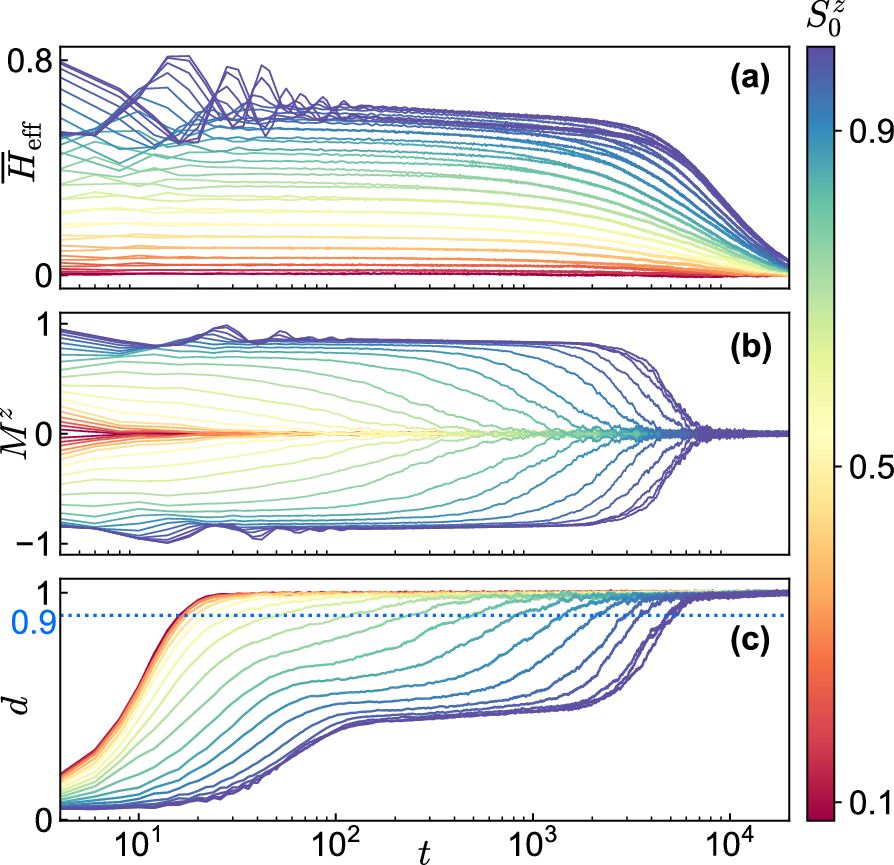}
    \caption{Initial energy-density dependence of {\color{black}DTCs} in 1D
      classical systems with NN interactions. (a--c) Time evolutions of
      the mean values of the energy density
      $\overline{H}_{\textrm{eff}}$ (a), the magnetization ${M}^z$ (b),
      and the normalized decorrelator $d$ (c) for different initial
    states, with $S_0^z$ denoting the $z$-component value of the spin vector.}
    \label{fig:inis-Hmd}
  \end{figure}

  Moreover, to verify the robustness of DTC order with an imperfect
  global $x$-flip operation with a small error $\delta_r=0.03$~rad, we
  show the Fourier transform (FT) signal of the $M^z$ as a function
  of the oscillation frequency $\omega_k$
  in Fig.~\ref{fig:DFT}(c), which is compared to trivial rotations
  along the $x$-axis without interactions, as shown in
  Fig.~\ref{fig:DFT}(a) and Fig.~\ref{fig:DFT}(b), for the perfect
  ($\delta_r=0$~rad) and imperfect ($\delta_r=0.03$~rad) global
  $x$-flip operations, respectively.
  It shows that with a small error $\delta_r$ being added to the
  rotation angle $\pi$ per period, the
  trivial imperfect rotation leads to two peaks in the FT signals,
  which indicates the shift of angular
  frequency and the absence of DTC order. For the Floquet system
  described with Eq.~(\ref{eq1}) even with imperfect global $x$-flip
  operation, the FT signals of $M^z$ exhibit subharmonic peak at the frequency
  $\omega_\textrm{peak}=\pi$~\cite{Choi2017}. To quantitatively check
  the robustness of DTC order, we calculate the crystalline fraction:
  \begin{equation}
    f = \frac {|S(\omega_\textrm{peak})|^2}{\sum_k |S(\omega_k)|^2},
  \end{equation}
  which is defined as the ratio of the peak intensity to the total
  spectral intensity~\cite{Choi2017}. Here, the $S(\omega_k)$ is
  the FT signal as a function of the frequency $\omega_k$, and
  $\omega_\textrm{peak}$ is the peak frequency.
  As shown in Fig.~\ref{fig:DFT}(d), the crystalline fraction $f$
  stays close to $1$ for $\delta_r/\pi \in [-0.1, 0.27]$, which
  shows the robustness of PDTC order with the subharmonic response in
  the  Floquet driving system (\ref{eq1}).
  Note that the asymmetry of crystalline fraction with respect to the
  zero frequency in Fig.~\ref{fig:DFT}(d)
  results from the existence of longitudinal field along $x$-axis in $H_x$.

  Prethermalization and MBL are two key mechanisms for stabilizing DTC order.
  For MBL-induced DTCs, time crystalline order is independent of the
  choice of the initial states, and the
  symmetry- and ergodicity-breaking behavior persists to an
  arbitrarily late time~\cite{Yao2014,DeRoeck2017,
  Nurwantoro2019, Kyprianidis2021,Liu2023}. In comparison,
  {the thermalization time of PDTCs grows exponentially with the
    increase of the driving frequency
    and depends on the energy density of the initial
  state}~\cite{Else2017,Machado2020,Kyprianidis2021}.
  Thus, to distinguish the {\color{black}prethermalization-like DTCs}
  from the MBL-induced DTCs, we investigate the thermalization time
  $\tau^*$ for different driving frequencies, which are shown in
  Fig.~\ref{fig:1dNN-dmH}(d).
  We observe that the Floquet thermalization time grows exponentially
  as the driving frequency increases in the parameter range
  $\omega/\pi\in[0.75,1.45]$.
  In addition, we show that the thermalization time $\tau^*$ depends on
  the energy density of the initial states. As shown in
  Fig.~\ref{fig:inis-Hmd}, dynamics of the mean values of the
  energy density $\overline{H}_\mathrm{eff}$, the magnetization
  $M^z$, and the normalized decorrelator $d$
  starting from different initial states have different prethermal
  plateaux and different thermalization times
  $\tau^*$. For the low-energy initial states, {\color{black}DTCs} do not exist,
  while {\color{black}DTC} order can be stabilized for
  high-energy initial states, which demonstrates the weak violation of ETH.
  Therefore, with the demonstration of {\color{black}two key features of
  PDTCs}, we have demonstrated the emergence of robust
  {\color{black}DTC order} in the 1D classical Floquet spin
  system with NN interactions.

  {\color{black}
  \section{Discussions on dynamics of the effective Hamiltonian and saturation of the thermalization time}

  \begin{figure}[t]
    \includegraphics[width=0.48\textwidth]{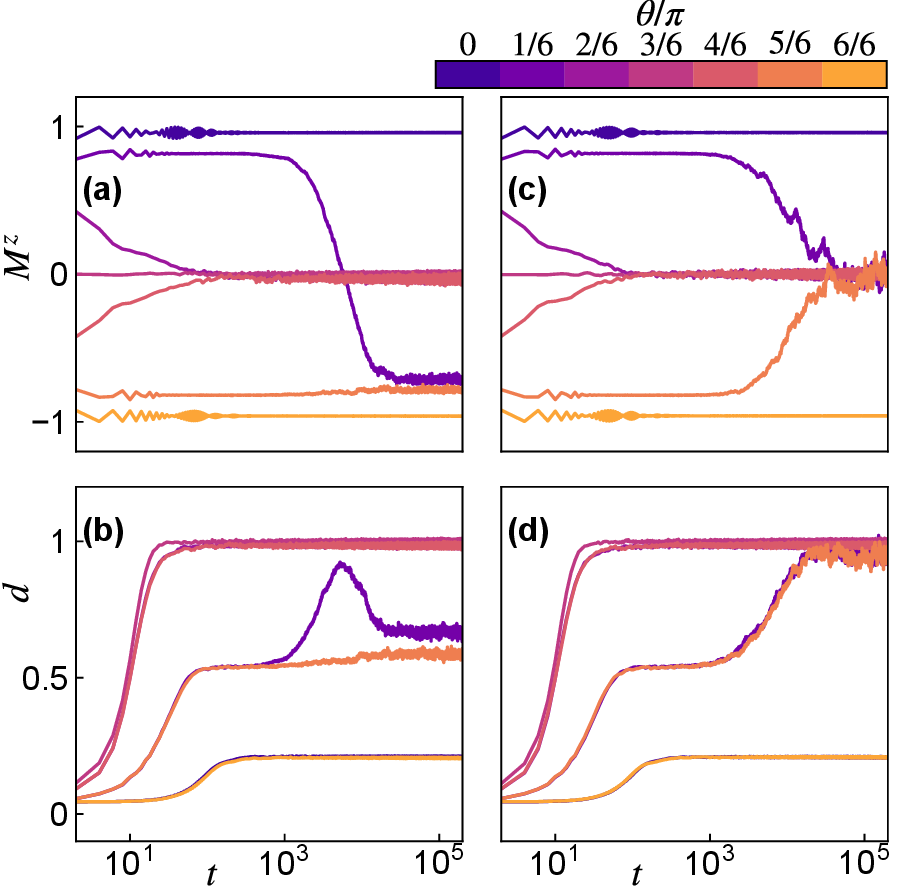}
    \caption{Time evolutions under the impact of $D^{(0)}$ and $D_x$ starting
      with different initial states with different initial average polar
      angles $\theta$.
      {\color{black}
      (a,b) Dynamics of the mean values of the magnetization $M^z$ (a) and
      the normalized decorrelator $d$ (b) under the impact of $D^{(0)}$.
      There exists flip of the magnetization for $\theta=\pi/6$.
      (c,d) Long-time behaviors of the magnetization $M^z$ (c) and the normalized
      decorrelator $d$ (d) in the toggling frame, during the dynamics under
      the impact of spin-flip-symmetric $D_x$. No magnetization flip is
      observed for  dynamics with $D_x$.
      Here, the $\theta / \pi$ is chosen as $\pm \frac{1}{6}$,
      $\pm \frac{2}{6}$, $\pm \frac{3}{6}$, and $0$. Results are obtained
      from the dynamics starting with 100 random initial states.
    }
  }
    \label{fig:Dpreth-Hmd}
  \end{figure}
  \begin{figure}[t]
    \includegraphics[width=0.48\textwidth]{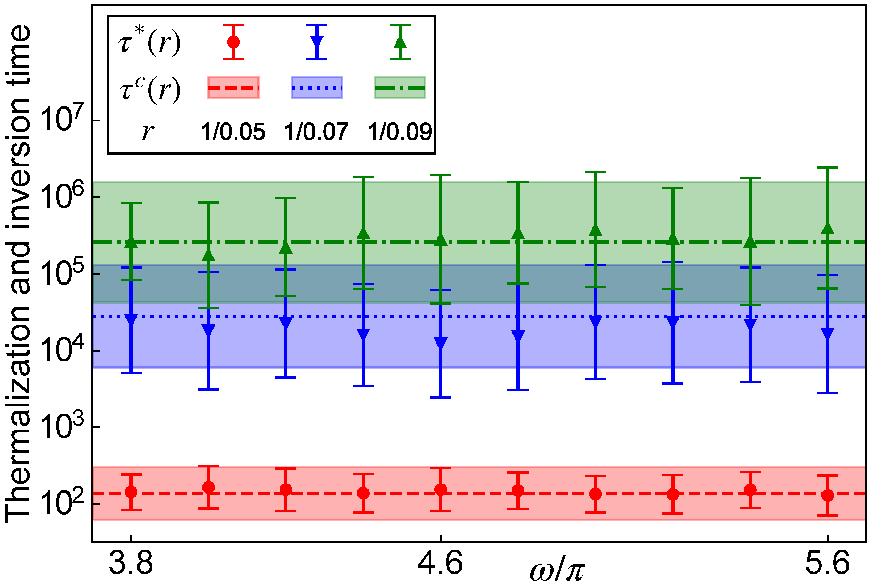}
    \caption{Saturation of the {\color{black}DTC} lifetimes with the increase
        of the driving frequency $\omega\in[3.8\pi, 5.6\pi]$. Here,
        $\tau^*(r)$ denotes the {\color{black}prethermalization-like DTC}
        lifetime, when $J_x$ and $b_z$ are simultaneously magnified by
        a ratio $r\in\{1/0.05, 1/0.07, 1/0.09\}$. Notation
        ${\tau}^{c}(r)$ denote the time {\color{black}when magnetization
        approaches zero during the dynamics under the impact of $D_x$.} Mean
        values of ${\tau}^{c}(r)$ with shaded areas denoting 1SD are obtained
        when $J_x$ and $b_z$ are
        magnified by $r\in\{1/0.05, 1/0.07, 1/0.09\}$. The initial
        mean value of the polar angles is chosen as $9\pi/10$, with
        other parameters being kept the same as previously listed. Results
        are obtained from the dynamics starting with 50 random initial states.}
    \label{fig:saturation_tauDX}
  \end{figure}

  For the Floquet Hamiltonian $H(t)$, we take the zeroth order of the
  truncated Floquet-Magnus expansion~\cite{Howell2019}
  \begin{align}
    D^{(0)}=\sum_{i=1}^N
    (2 J_z S_{i}^z S_{i+1}^z + 2 J_x S_{i}^x S_{i+1}^x  + b_z S_{i}^z
    + b_x S_{i}^x),
  \end{align}
  as the effective Hamiltonian.
  In the slightly rotated toggling frame, dynamics in one period of the Floquet system
  with spin flips $R_x(\pi)$ can be approximated as $R_x(\pi)\exp(-i D_x T)$,
  where
  \begin{equation}
  D_x = \sum_{i=1}^N
    (2 J_z S_{i}^z S_{i+1}^z + 2 J_x S_{i}^x S_{i+1}^x + b_x S_{i}^x)
  \end{equation}
  denotes the terms that commute with $R_x(\pi)$ in $D^{(0)}$~\cite{Ye2021}.
  Both $D^{(0)}$ and $D_x$ are expected to thermalize rapidly in
  prethermalization and PDTCs, respectively~\cite{Else2017}.

  However, when we consider the dynamics generated by $D^{(0)}$ and $D_x$, the
  time evolutions starting from the initial states with high energy densities
  is hard to thermalize, showing a violation of ETH, as shown in
  Fig.~\ref{fig:Dpreth-Hmd}. For the slow thermalization behavior
  of the effective Hamiltonian $D_x$, our results might not comply
  with the conventional picture of prethermalization and PDTCs.
  The DTCs in our model may result from the slow thermalizing behavior in
  the initial-state-persistent dynamics of $D_x$, and the DTC lifetimes are
  finally constrained by the thermalization behavior of $D_x$ in the high
  driving frequency limit, as shown with the saturation of the thermalization time.
    This can be observed in the high frequency regime, as shown
    in Fig.~\ref{fig:saturation_tauDX}. 

    In addition, we find the saturation DTC lifetimes coincide with
    $\tau^c$ in Fig.~\ref{fig:saturation_tauDX},
    which is defined as the time when $M^z$ approaches zero during the
    time evolution with $D_x$ in the toggling frame. It can be inferred
    that before the saturation,
    the DTC lifetime is bounded by the operator error, which is
    exponentially small with the increase of driving frequency
    as depicted in the conventional picture of PDTCs, and the saturation
    thermalization time of DTCs is bounded by $\tau^c$ that characterizes
    the thermalization behavior of $D_x$.
    Moreover, when $J_x$ and $b_z$ are magnified by the same
    ratio $r\in\{1/0.05, 1/0.07, 1/0.09\}$, we observe the
    exponential increase of the thermalization time as $J_x$ and $b_z$
    decrease simultaneously.
    This is similar to prethermalization that is caused by
    integrability-breaking as introduced in Ref.~\cite{Chen2023}.
    It can also be inferred that the long thermalization time of $D_x$,
    for the evolution from the initial state without
    spin-flip symmetry to the spin-flip-symmetric Gibbs state, makes the
    subharmonic response observable.
    Therefore, the mechanism for the DTCs in our model would not
    comply with the conventional picture of PDTCs as introduced in
    Refs.~\cite{Else2017, Ye2021}, which would be a topic for
    further investigations.
}

  \section{conclusion}

  In summary, we demonstrate the emergence of DTC order in disorder-free
  1D classical Floquet spin systems with only NN interactions. We observe
  DTC order {\color{black}first} with an exponential growth of the thermalization time
  as the driving frequency increases, which then reaches saturation.
  The lifetime of the DTCs in 1D classical systems, characterized by the
  thermalization time, also depends on the energy density of the initial states.
  Therefore, we verify the existence of {\color{black}prethermalization-like
  DTCs, since the rapid thermalization of the effective Hamiltonian
  is absent.} The robustness of DTC order is investigated by introducing
  imperfect $x$-flip operations. Our work shows the existence of
  {\color{black}DTCs} in classical Floquet spin
  systems with only 1D NN interactions, and extends the understandings of
  DTCs in disorder-free classical spin
  systems~\cite{Pizzi2021,Pizzi2021a, Ye2021}.
  The DTCs in 1D classical systems could be probed in several experimental
  platforms, including LC circuits and mechanical
  gyroscopes~\cite{SerraGarcia2019,Wu2020,Li2018,Wang2020}.
  The DTCs in 1D classical systems with only NN interactions could
  not be well explained using previous theories proposed in
  Refs.~\cite{Else2017, Ye2021}, {\color{black}which would be a topic
  for further investigations}. The intriguing properties demonstrated with
  classical DTC order might also inspire further advancements in the study
  of corresponding quantum models.

  \begin{acknowledgments}
    This work is partially supported by the National Natural Science
    Foundation of China (Grant No.~12475017) and the
    Natural Science Foundation of Guangdong Province (Grant
    No.~2024A1515010398),
    and the Startup Grant of South
    China University of Technology (Grant No.~20240061).

  \end{acknowledgments}

  \appendix

  \section{\label{app.unitary}Demonstration of unchanged length of
  classical spins in evolution}

  Here, we study the dynamics of the 1D classical spin chain with the
  vector length for the $i$-th spin
  {\color{black}being initially unit: $|\mathbf{S}_i(0)|=1$}.
  For simplicity, we first consider the time evolution for the first
  half period with $t\,\textrm{mod}\,2T\in[0,T]$, under the impact of
  the non-flipping Hamiltonian $H_z=\sum_i (S_i^zS_{i+1}^z+S_i^z)$ in
  Eq.~\eqref{eq2},
  where all coefficients, without loss of generality, are set to $1$.
  {\color{black}To verify $|\mathbf{S}_i(t)|=1$, we need to prove
    \begin{align}
      \label{App-req}
      \frac{\partial }{\partial t} |\mathbf{S}_i(t)|^2 = 0
    \end{align}
    with $|\mathbf{S}_i(0)|=1$. Equation~\ref{App-req} can be written as
    \begin{align}
      \mathbf{v}_1 \cdot \mathbf{v}_2  = 0
    \end{align}
    where we have defined that $\mathbf{v}_1\equiv(S_i^x,S_i^y,S_i^z)$, and
    $\mathbf{v}_2\equiv(\{S_i^x,H\},\{S_i^y,H\},\{S_i^z,H\})$.
  }
  {\color{black} In our model, with}
  \begin{align}
    \{S_i^\alpha, \sum_i S_i^zS_{i+1}^z\} &=
    -\epsilon_{\alpha\beta\gamma} S_{i}^\beta (S_{i-1}^\gamma + S_{i+1}^\gamma),
  \end{align}
  and
  \begin{align}
    \{S_i^\alpha, \sum_j S_j^\beta\} &= \epsilon_{\alpha\beta\gamma} S^\gamma,
  \end{align}

  \begin{figure}[b]
    \includegraphics[width=0.48\textwidth]{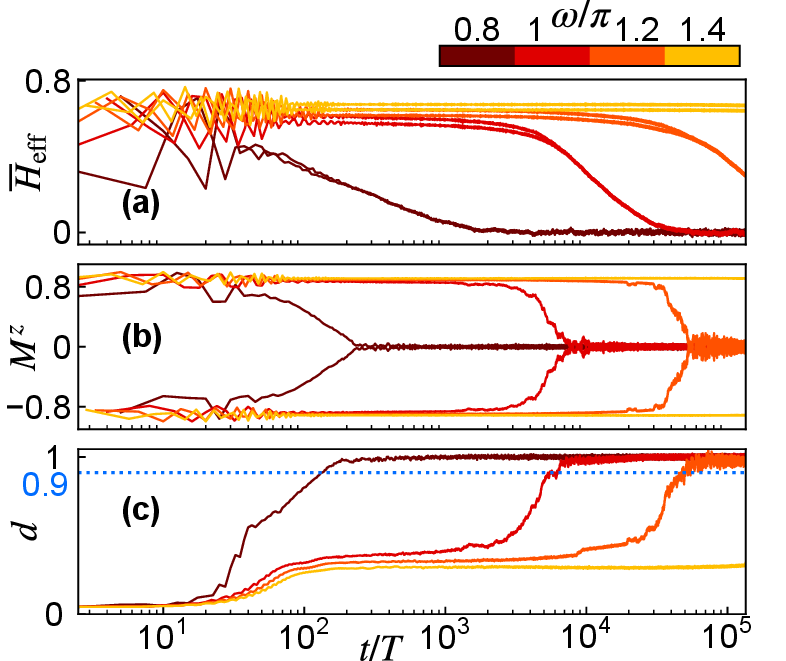}
    \caption{{\color{black}Emergence of DTCs for another set of parameters.
        (a--c) Dynamics of the average energy density
        $\overline{H}_\mathrm{eff}$ (a), the
        magnetization along $z$-direction $M_z$ (b), and the
        normalized decorrelator $d$ (c) show the
        existence of DTCs. Mean values of the three observables are
        calculated from 30 random initial
        states at the stroboscopic times $t = mT$, with $m \in
        \mathbb{N}$. Parameters are chosen as
    $J_z=0.36, J_x=0.01, b_z=-0.014$, and $b_x=-0.33$.}}
    \label{fig:Hmd_paras}
  \end{figure}

  the spin components evolve as
  \begin{align}
    \frac{\partial S_i^x}{\partial t} &= -S_i^y (1+S_{i-1}^z+S_{i+1}^z), \\
    \frac{\partial S_i^y}{\partial t} &= S_i^x (1+S_{i-1}^z+S_{i+1}^z), \\
    \frac{\partial S_i^z}{\partial t} &= 0.
  \end{align}
  {\color{black}It means that} during each first half period
  $t\,\textrm{mod}\,2T\in[0,T]$, the $z$-components of all spins are
  unchanged, and
  all spin  vectors are rotated along the positive $z$-axis as
  \begin{equation}
    \frac{\partial }{\partial t}\mathbf{S}_i = C \hat{z}\times \mathbf{S}_i,
  \end{equation}
  with a constant angular frequency $C=1+S_{i-1}^z+S_{i+1}^z$. Thus,
  we obtain that
  \begin{align}
    S_i^x\frac{\partial  S_i^x}{\partial t} +  S_i^y\frac{\partial
    S_i^y}{\partial t} +  S_i^z\frac{\partial  S_i^z}{\partial t} =0,
  \end{align}
  which {\color{black}closes the proof of Eq.~\ref{App-req}.
  Therefore, the equations of motion keeps} the unit of the classical
  spin vector length. 
  The time evolution during the second half period
  $t\,\textrm{mod}\,2T\in[T,2T]$ under the impact of $H_x=\sum_i
  (S_i^x S_{i+1}^x +S_i^x)$
  is similar to that of the first half period with $H_z$,
  by replacing $z\to x$.

    \section{\label{app:para_robust}
      Emergence of PDTCs in 1D classical Floquet systems with NN interactions
  for another set of parameters}

    To show that our results can be obtained from a range of
    parameters in the Hamiltonian, the existence
    of PDTCs with different parameters is shown in Fig.~\ref{fig:Hmd_paras}.

  \bibliography{v7_1dNN_xflip.bib}

  \end{document}